\documentclass[12pt,epsfig]{iopart}

\usepackage{axodraw}
\usepackage{graphicx}

\newcommand{\hs}{\hspace*{0.5cm}}

\newcommand{\be}{\begin{equation}}
\newcommand{\ee}{\end{equation}}
\newcommand{\bea}{\begin{eqnarray}}
\newcommand{\eea}{\end{eqnarray}}
\newcommand{\bary}{\begin{array}}
\newcommand{\eary}{\end{array}}
\newcommand{\bit}{\begin{itemize}}
\newcommand{\eit}{\end{itemize}}
\newcommand{\ben}{\begin{enumerate}}
\newcommand{\een}{\end{enumerate}}
\newcommand{\crn}{\nonumber \\}

\newcommand{\al}{\alpha}
\newcommand{\la}{\lambda}

\newcommand{\ga}{\gamma}
\newcommand{\va}{\varphi}

\newcommand{\fr}{\frac}

\newcommand{\bc}{\begin{center}}
\newcommand{\ec}{\end{center}}
\newcommand{\nn}{\nonumber}

\begin{document}

\title[]{ Non-thermal leptogenesis in
supersymmetric 3-3-1 model with inflationary scenario}

\author{ D. T. Huong$^{a,b}$ and
H. N. Long$^{a,c}$ }

\address{
$^a$\footnote{Permanent address} Institute of  Physics, VAST, P.
O. Box 429, Bo Ho, Hanoi 10000, Vietnam\\
$^b$  Theory Division, Institute of Particle and Nuclear Studies,
High Energy Accelerator Research Organization (KEK),
Tsukuba, Ibaraki 305-0801, Japan\\
 $^c$ Saha Institute of Nuclear Physics, 1/AF Bidhan-Nagar, Calcutta 700064,
 India} \ead{hnlong@iop.vast.ac.vn}
\begin{abstract}

We study a leptogenesis scenario in which the heavy Majorana
neutrinos are produced  non-thermally
 in inflaton decays in the supersymmetric
economical  $\mathrm{SU}(3)_C\otimes \mathrm{SU}(3)_L \otimes
\mathrm{U}(1)_X$  model with inflationary scenario, and for this
purpose neutrino masses play the key role. Due to the  inflaton
with mass in the GUT scale, the  model under consideration
provides successful neutrino masses, which is different from ones
without inflationary scenario.  The lepton-number-violating
interactions among the inflaton and right-handed neutrinos appear
at the one-loop level, and this is a reason for non-thermal
leptogenesis scenario. The bound followed from the gravitino
abundance and the cosmological
 constraint on  neutrino mass/the neutrino oscillation data is
  $m_{\nu 3} \simeq \frac{0.05}{\delta_{eff}} $
 eV. By taking the reheating temperature as low as $T_R= 10^6$
 GeV, we get a limit on the ratio of masses of the light
 heavies neutrino to those of the inflaton to be  $
\fr{M_{R1}}{M_{\phi}} = 0.87$.

\end{abstract}
\begin{itemize}
\item PACS: 98.80.Cq, 12.60.Jv, 14.60.St
\end{itemize}

 \maketitle

\section{Introduction}

The recent experimental results confirm that neutrinos have tiny
masses and oscillate \cite{pdg}, which implies that the standard
model (SM) must be extended. Among the beyond-SM extensions, the
models based on the $\mathrm{SU}(3)_C\otimes \mathrm{SU}(3)_L
\otimes \mathrm{U}(1)_X$ (3-3-1) gauge group \cite{ppf,flt} have
some intriguing features. First, they can give partial explanation
of the generation number problem. Second,  the third quark
generation has to be different from the first two, so this leads
to the possible explanation of why top quark is
uncharacteristically heavy. An additional motivation to study this
kind of the models is that they can also predict the electric
charge quantization \cite{prs}.

Depending on the electric charge of particle at the bottom of the
lepton triplet, the 3-3-1 models are classified into two main
versions: the minimal model \cite{ppf} with the lepton triplet
$\left(\nu, l, l^c\right)_L$ and the version with right-handed
(RH) neutrinos  \cite{flt}, where the RH neutrinos place at the
bottom of the triplet: $\left(\nu, l, \nu^c\right)_L$.  In the
3-3-1 model with right-handed neutrinos, the scalar sector
requires three Higgs triplets. It is interesting to note that two
Higgs triplets of this model have the same $\mathrm{U}(1)_X$
charges with two neutral components at their top and bottom. In
the model under consideration, the new charge $X$ is connected
with the electric charge operator through a relation \be
Q=T_3-\fr{1}{\sqrt{3}}T_8+X.\ee Assigning these neutral components
vacuum expectation values (VEVs) we can reduce the number of Higgs
triplets to  two. Therefore we have a resulting 3-3-1 model with
two Higgs triplets \cite{haihiggs}. As a result, the dynamical
symmetry breaking also affects the lepton number. Hence it follows
that the lepton number is also broken spontaneously at a high
scale of energy. Note that the mentioned model contains a very
important advantage, namely, there is no new parameter, but it
contains very simple Higgs sector; therefore, the significant
number of free parameters is reduced. To mark the minimal content
of the Higgs sector, this version that includes right-handed
neutrinos is going to be called the {\it economical 3-3-1 model}.

By this time, the  cosmology  becomes one of the most important
sciences giving deep knowledge on the origin of our Universe. The
critical moment for the development of modern cosmology was
discovery of the 2.7 $K$ microwave background radiation arriving
from the farthest reaches of the universe. The existence of the
microwave background had been predicted by the hot-universe
theory, which gained immediate and widespread acceptance after the
discovery. Despite successes,  there are a lot of difficulties
(see, for example,~\cite{linde1}) in modern cosmology such as
flatness, horizon, primordial monopole problems,  etc. It is all
the more surprising, then, that many of these problems, together
with a number of others that predate the hot universe theory, have
been resolved in the context of one fairly simple scenario for the
development of the universe - the so-called inflationary universe
scenario~\cite{infsce}. Inflation assumes that there was a period
in the very early universe when the potential  and vacuum energy
density  dominated the energy of the universe, so that the cosmic
scale factor grew exponentially.
  The important ingredient of
the inflationary  scenario is a scalar field $\va$ having
effective potential $V(\va)$ with some properties (satisfying many
constrains that are rather unnatural). This scalar field is called
inflaton.

On the other hand, to explain the well-known matter-antimatter
asymmetry, the baryogenesis  plays an important role. In addition,
primordial lepton asymmetry is converted to baryon asymmetry in
the early universe through the ``sphaleron'' effects of
electroweak gauge theories \cite{spha} if it is produced before
the electroweak phase transition.  Thus, the leptogenesis scenario
\cite{leptogenesis} seems to be the most plausible mechanism for
creating the cosmological baryon asymmetry.

The aim of this work is to consider leptogenesis in the
supersymmetric economical 3-3-1 model with inflationary scenario.
This paper is organized as follows. In Section ~\ref{parcontent}
we present the particle  content in the supersymmetric economical
3-3-1 model. Section \ref{fmass} is devoted to neutrino mass in
the  supersymmetric economical 3-3-1 model without the
inflationary scenario. We will show that in this case, the
neutrino mass matrix is  unrealistic. In Section~\ref{seesaw}, we
present the seesaw mechanism in the model with inflationary
scenario. At the one-loop level, the neutrino mass matrix gives
the necessary hierarchy. The non-thermal leptogenesis scenario is
presented in section \ref{nontermal}.  We summary our results and
make conclusions in section~\ref{conclusion}.

\section{\label{parcontent}A brief review of the model}

To proceed further, the necessary features of the supersymmetric
economical 3-3-1 model \cite{susyeco,higph} will be presented. The
superfield content in this model is defined in a standard way as
follows: \be \widehat{F}= (\widetilde{F}, F),\hs \widehat{S} = (S,
\widetilde{S}),\hs \widehat{V}= (\lambda,V), \ee where the
components $F$, $S$ and $V$ stand for the fermion, scalar and
vector fields of the economical 3-3-1 model while their
superpartners are denoted as $\widetilde{F}$, $\widetilde{S}$ and
$\lambda$, respectively \cite{susy,s331r}.

The superfields for the leptons under the 3-3-1 gauge group
transform as
\begin{equation}
\widehat{L}_{a L}=\left(\widehat{\nu}_{a}, \widehat{l}_{a},
\widehat{\nu}^c_{a}\right)^T_{L} \sim ( {\bf 1},{\bf 3},-1/3),\hs
  \widehat {l}^{c}_{a L} \sim ({\bf 1},{\bf 1},1),\label{l2}
\end{equation} where $\widehat{\nu}^c_L=(\widehat{\nu}_R)^c$ and $a=1,2,3$
is a generation index.

It is worth mentioning that in the economical version the first
generation of quarks should be different from others
\cite{haihiggs}. The superfields for the left-handed quarks of the
first generation are in triplets \be \widehat Q_{1L}=
\left(\widehat { u}_1,\
                        \widehat {d}_1,\
                        \widehat {u}^\prime
 \right)^T_L \sim ( {\bf 3},{\bf 3},1/3),\label{quarks3}\ee
where the right-handed singlet counterparts are given by\be
\widehat {u}^{c}_{1L},\ \widehat { u}^{ \prime c}_{L} \sim ( {\bf
3^{*}},{\bf 1},-2/3),\hs \widehat {d}^{c}_{1L} \sim ( {\bf
3^{*}},{\bf 1},1/3 ). \label{l5} \ee Conversely, the superfields
for the last two generations transform as antitriplets
\begin{equation}
\begin{array}{ccc}
 \widehat{Q}_{\alpha L} = \left(\widehat{d}_{\alpha}, - \widehat{u}_{\alpha},
 \widehat{d^\prime}_{\alpha}\right)^T_{L} \sim (
 {\bf 3},{\bf 3^*},0), \hs \al=2,3, \label{l3}
\end{array}
\end{equation}
where the right-handed counterparts are in singlets
\begin{equation}
\widehat{u}^{c}_{\alpha L} \sim \left( {\bf 3^{*}},{\bf 1},-2/3
\right),\hs \widehat{d}^{c}_{\alpha L},\ \widehat{d}^{\prime
c}_{\alpha L} \sim \left( {\bf 3^{*}},{\bf 1},1/3 \right).
\label{l4}
\end{equation}

The primes superscript on usual quark types ($u'$ with the
electric charge $q_{u'}=2/3$ and $d'$ with $q_{d'}=-1/3$) indicate
that those quarks are exotic ones. The mentioned fermion content,
which belongs to that of the 3-3-1 model with right-handed
neutrinos \cite{flt,haihiggs} is, of course,  free from anomaly.

The two superfields $\widehat{\chi}$ and $\widehat {\rho} $ are
introduced to span the scalar sector of the economical 3-3-1
model \cite{haihiggs}: \bea \widehat{\chi}&=& \left (
\widehat{\chi}^0_1, \widehat{\chi}^-, \widehat{\chi}^0_2
\right)^T\sim (1,3,-1/3), \label{l7}\\
\widehat{\rho}&=& \left (\widehat{\rho}^+_1, \widehat{\rho}^0,
\widehat{\rho}^+_2\right)^T \sim  (1,3,2/3). \label{l8} \eea To
cancel the chiral anomalies of Higgsino sector, the two extra
superfields $\widehat{\chi}^\prime$ and $\widehat {\rho}^\prime $
must be added as follows: \bea \widehat{\chi}^\prime&=& \left
(\widehat{\chi}^{\prime 0}_1, \widehat{\chi}^{\prime
+},\widehat{\chi}^{\prime 0}_2 \right)^T\sim ( 1,3^*,1/3),
\label{l9}\\
\widehat{\rho}^\prime &=& \left (\widehat{\rho}^{\prime -}_1,
  \widehat{\rho}^{\prime 0},  \widehat{\rho}^{\prime -}_2
\right)^T\sim (1,3^*,-2/3). \label{l10} \eea

In this model, the $ \mathrm{SU}(3)_L \otimes \mathrm{U}(1)_X$
gauge group is broken via two steps:
 \be \mathrm{SU}(3)_L \otimes
\mathrm{U}(1)_X \stackrel{w,w'}{\longrightarrow}\ \mathrm{SU}(2)_L
\otimes \mathrm{U}(1)_Y\stackrel{v,v',u,u'}{\longrightarrow}
\mathrm{U}(1)_{Q},\label{stages}\ee where the VEVs are defined by
\bea
 \sqrt{2} \langle\chi\rangle^T &=& \left(u, 0, w\right), \hs \sqrt{2}
 \langle\chi^\prime\rangle^T = \left(u^\prime,  0,
 w^\prime\right),\\
\sqrt{2}  \langle\rho\rangle^T &=& \left( 0, v, 0 \right), \hs
\sqrt{2} \langle\rho^\prime\rangle^T = \left( 0, v^\prime,  0
\right).\eea The VEVs $w$ and $w^\prime$ are responsible for the
first step of the symmetry breaking while $u,\ u^\prime$ and $v,\
v^\prime$ are for the second one. The VEVs $w, w^\prime$ give mass
for the exotic quarks and new gauge bosons while the VEVs $u,
u^\prime, v, v^\prime $ give mass for SM particles. Therefore they
have to satisfy the constraints
 \be
 u,\ u^\prime,\ v,\ v^\prime
\ll w,\ w^\prime. \label{contraint}\ee
 On the other hand, we can
drive  constraint $v, v^\prime \simeq v_{electroweak}=246$ GeV
from the bound of $W$ boson mass and $u, u^\prime < 2.46 $ GeV
(for details, see \cite{haihiggs}). Note that $u$ and $u^\prime$
carry lepton number 2 \cite{changlong}, so they are the kinds of
lepton-number-violating parameter. Hence, it leads to the limit
\be u, u^\prime \ll v, v^\prime.\label{coduu} \ee

 The vector
superfields $\widehat{V}_c$, $\widehat{V}$ and
$\widehat{V}^\prime$ containing the usual gauge bosons are,
respectively, associated with the $\mathrm{SU}(3)_C$,
$\mathrm{SU}(3)_L$ and $\mathrm{U}(1)_X $ group factors. The
colour and flavour vector superfields have expansions in the
Gell-Mann matrix bases $T^d=\lambda^d/2$ $(d=1,2,...,8)$ as
follows:\bea \widehat{V}_c &=& \fr{1}{2}\lambda^d
\widehat{V}_{cd},\hs
\widehat{\overline{V}}_c=-\fr{1}{2}\lambda^{d*}
\widehat{V}_{cd};\hs \widehat{V} = \fr{1}{2}\lambda^d
\widehat{V}_{d},\hs \widehat{\overline{V}}=-\fr{1}{2}\lambda^{d*}
\widehat{V}_{d}, \crn \eea where an overbar $^-$ indicates complex
conjugation. For the vector superfield associated with
$\mathrm{U}(1)_X$, we normalize as follows \be X \hat{V}'= (XT^9)
\hat{B}, \hs T^9\equiv\fr{1}{\sqrt{6}}\mathrm{diag}(1,1,1).\ee In
the following, we  denote the gluons by $g^d$ and their respective
gluino partners by $\lambda^d_{c}$, with $d=1, \ldots,8$. In the
electroweak sector, $V^d$ and $B$ stand for the
$\mathrm{SU}(3)_{L}$ and $\mathrm{U}(1)_{X}$ gauge bosons with
their gaugino partners $\lambda^d_{V}$ and $\lambda_{B}$,
respectively.

 \hs The most general superpotential is obtained by \cite{susyeco,jhep}
 \be
 W=W_2+W_3
 \ee
with
\begin{eqnarray}
W_{2}&=& \mu_{0a}\hat{L}_{aL} \hat{ \chi}^{\prime}+ \mu_{ \chi}
\hat{ \chi} \hat{ \chi}^{\prime}+
 \mu_{ \rho} \hat{ \rho} \hat{ \rho}^{\prime},
 \label{w2}
\end{eqnarray}
and

\begin{eqnarray}
W_{3}&=& \ga_{ab} \hat{L}_{aL} \hat{ \rho}^{\prime}
\hat{l}^{c}_{bL}+ \la_{a} \epsilon \hat{L}_{aL} \hat{\chi}
\hat{\rho}+ \la^\prime_{ab} \epsilon \hat{L}_{aL} \hat{L}_{bL}
\hat{\rho} \nonumber \\
&+& \kappa_{i} \hat{Q}_{1L} \hat{\chi}^{\prime} \hat{u}^{c}_{iL}+
\kappa^\prime \hat{Q}_{1L} \hat{\chi}^{\prime} \hat{u}^{\prime
c}_L+ \vartheta_{i}\hat{Q}_{1L} \hat{\rho}^{\prime}
\hat{d}^{c}_{iL} \nonumber \\
&+& \vartheta^\prime_{ \alpha}\hat{Q}_{1L} \hat{\rho}^{\prime}
\hat{d}^{\prime c}_{\alpha L} + \pi_{ \alpha i} \hat{Q}_{\alpha
L}\hat{\rho}\hat{u}^{c}_{iL} +\pi_{\alpha}^{\prime}
\hat{Q}_{\alpha L}\hat{\rho}\hat{u}^{\prime c}_{L} \nonumber \\
&+& \Pi_{\alpha i} \hat{Q}_{\alpha L} \hat{\chi} \hat{d}^{c}_{iL}
+ \Pi^\prime_{\alpha \beta} \hat{Q}_{\alpha L} \hat{\chi}
\hat{d}^{\prime c}_{\beta L}+ \epsilon
f_{\alpha\beta\gamma}\hat{Q}_{\alpha L} \hat{Q}_{\beta L}
\hat{Q}_{\gamma L} \crn &+& \xi_{1i \beta j} \hat{d}^{c}_{iL}
\hat{d}^{\prime c}_{\beta L} \hat{u}^{c}_{j L}+ \xi_{2i \beta }
\hat{d}^{c}_{i L} \hat{d}^{\prime c}_{\beta L} \hat{u}^{\prime
c}_{L}+ \xi_{3ijk}
\hat{d}^{c}_{iL} \hat{d}^{c}_{jL} \hat{u}^{c}_{k L} \nonumber \\
&+& \xi_{4ij} \hat{d}^{c}_{i L} \hat{d}^{c}_{jL} \hat{u}^{\prime
c}_{L}+ \xi_{5 \alpha \beta i} \hat{d}^{\prime c}_{\alpha L}
\hat{d}^{\prime c}_{\beta L} \hat{u}^{c}_{iL} + \xi_{6 \alpha
\beta} \hat{d}^{\prime c}_{\alpha L}\hat{d}^{\prime c}_{\beta L}
\hat{u}^{\prime c}_{L} \crn &+& \xi_{a \alpha j}\hat{L}_{aL}
\hat{Q}_{\alpha L} \hat{d}^{c}_{jL}+ \xi^\prime_{a\alpha
\beta}\hat{L}_{aL} \hat{Q}_{\alpha L} \hat{d}^{\prime c}_{\beta
L}. \label{w3}
\end{eqnarray}
where  $a=1,2,3$,  $i=1,2,3$ and $\al, \beta =2,3$.  The
coefficients $\mu_{0a}, \mu_{\rho}$ and $\mu_{\chi}$ have mass
dimension, while all  coefficients in $W_{3}$ are dimensionless
and $\la^\prime_{ab}= - \la^\prime_{ba}$.

\section{\label{fmass} Neutrino mass in supersymmetric 3-3-1
 model without the inflationary scenario}

Let us mention that recent data from neutrino oscillations
produced the following results:
 \bea 0.36 \leq \sin^2 \theta_{23}
\leq 0.67,\hs 0.27 \leq \sin^2 \theta_{12} \leq  0.38 \hs \sin^2
\theta_{13} < 0.053,\label{ndata1} \eea and \bea 2.07 \times
10^{-3}\ \textrm{eV}^2 & \leq & \Delta m^2_{atm}  \leq 2.75 \times
10^{-3}\
\textrm{eV}^2,\label{ndata2}\\
 7.03 \times 10^{-5}\
\textrm{eV}^2 &\leq & \Delta m^2_{sol}  \leq 8.27 \times 10^{-5}\
\textrm{eV}^2 ,\nn \eea at 99.73\% CL \cite{neudata}.

This gives the constraints on neutrino masses and mixing. Let us
consider the above problem in the model the without inflationary
scenario.

\subsection{Tree-level Dirac mass}

At the tree-level, the neutrinos get masses from the term
\begin{equation}
-\lambda_{ab}^\prime L_{aL}L_{bL}\rho+ H.c,
\end{equation}
which gives us
\begin{equation}
-\la_{ab}^\prime ( \nu^{c}_{aL}\nu_{bL}- \nu_{aL}\nu^{c}_{bL}+
\overline{\nu^{c}_{aL}}\overline{\nu_{bL}}-
\overline{\nu_{aL}}\overline{\nu^{c}_{bL}}) \rho^{0}.
\end{equation}
This mass term can now be rewritten in terms of a $6 \times 6$
matrix $X_{\nu}$ by defining the following column vector:
\begin{equation}
( \psi^{0}_{\nu})^{T}= \left( \begin{array}{cccccc} \nu_{1L} &
\nu_{2L} & \nu_{3L} & \nu^{c}_{1L} & \nu^{c}_{2L} & \nu^{c}_{3L}
\end{array} \right) .
\end{equation}
Now we can rewrite our mass term as
\begin{equation}
- {\cal L} =\fr 1 2 \left[
(\psi^{0}_{\nu})^{T}X_{\nu}\psi^{0}_{\nu}+ H.c \right],
\end{equation}
with
\begin{equation}
X_{\nu}= \frac{v}{\sqrt{2}}\left(
\begin{array}{cccccc}
  0 & 0 & 0 & 0 & G_{21} & G_{31} \\
  0 & 0 & 0 & G_{12} & 0 & G_{32} \\
  0 & 0 & 0 & G_{13} & G_{23} & 0 \\
  0 & G_{12} & G_{13} & 0 & 0 & 0 \\
  G_{21} & 0 & G_{23} & 0 & 0 & 0 \\
  G_{31} & G_{32} & 0 & 0 & 0 & 0 \\
\end{array}
\right) \nn \equiv  \left(
                 \begin{array}{cc}
                   0 & M_D^T \\
                   M_D & 0\\
                 \end{array}
               \right)
\end{equation}
where
\begin{equation}
G_{ab}= \left( \la_{ab}^\prime - \lambda_{ba}^\prime \right).
\end{equation}
 Due to the
fact that $G_{ab}=-G_{ba}$, the mass pattern of this sector is
$0,\ 0,$ $\ m_{\nu},\ m_{\nu},$ $\ m_{\nu},\ m_{\nu}$, where
$\sqrt{2}m_{\nu}=v\sqrt{G^{2}_{31}+G^{2}_{32}+G^{2}_{21}}$. Noting
that this mass spectrum is the same as that of the
non-supersymmetric version and the mass spectrum is not realistic
\cite{dls1}. The most general neutrino mass spectrum is in the
following form:
\bea M_{\nu}=\left(%
\begin{array}{cc}
  M_{L} & M_D^T \\
  M_D & M_{R} \\
\end{array}%
\right),\label{matran}\eea where $M_{L,R}$ (vanish at the
tree-level) and $M_D$ get possible corrections.

\subsection{The one-loop corrections to the Dirac and Majorana masses}
\label{oneloop} The Yukawa couplings of the leptons and the
relevant Higgs self-couplings are explicitly rewritten as follows:
\bea
L^{lept}_{Y}&=&\la^\prime_{ab}\nu_{aL}l_{bL}\rho^+_3+\la^\prime_{ab}\nu_{aR}^c
l_{bL}\rho^+_1+\ga_{ab}\nu_{aL}l_R^c\rho_1^{\prime -}
+\ga_{ab}\nu_{aR}^cl_R^c \rho_3^{\prime -} +H.c.,\crn
L^{relv}_{H}&=& \frac{g^2}{8}(\chi^\dagger_i\lambda^b_{ij}\chi_j-
\chi^{\prime\dagger}_i\lambda^{*b}_{ij}\chi^\prime_j+
\rho^\dagger_i\lambda^b_{ij}\rho_j-
\rho^{\prime\dagger}_i\lambda^{*b}_{ij}\rho^\prime_j)^2\crn &+&
\frac{g^{\prime2}}{12}
 \left(-\frac{1}{3} \chi^{\dagger} \chi +
\frac{1}{3} \chi^{\prime \dagger} \chi^{\prime}+ \frac{2}{3}
\rho^{\dagger}\rho - \frac{2}{3} \rho^{\prime
\dagger}\rho^{\prime} \right)^2\label{t9181} \eea In the limit
$v,v^\prime, u, u^\prime \ll w, w^\prime$, the masses of the
charged Higgs bosons get approximate values such as \cite{higph}:
$m_{\rho_1^{\prime -}}\simeq m_W, m_{\rho_1^+}\simeq 0,
m_{\rho_3^+}\simeq m_{\zeta_2}=0, m_{\rho_3^ {\prime -}}\simeq
m_{\zeta_3}=0$.

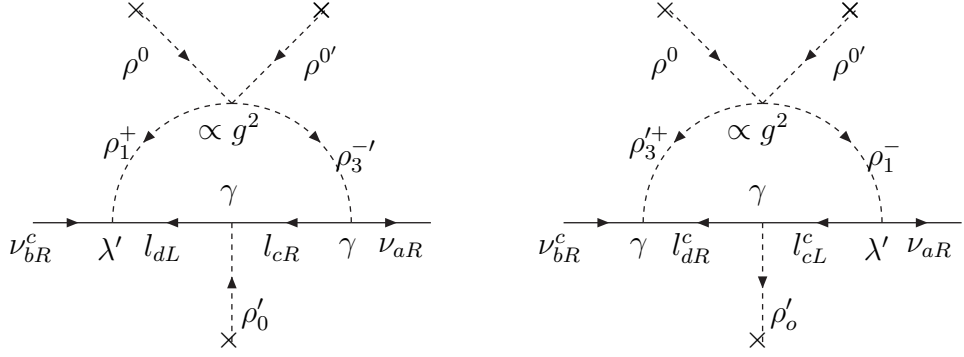
\begin{figure}[htbp]
\begin{center}
\begin{picture}(600,120)(150,-150)
\ArrowLine(210,-120)(240,-120) \ArrowLine(330,-120)(360,-120)
\ArrowLine(285,-120)(240,-120) \ArrowLine(330,-120)(285,-120)
\DashArrowLine (285,-165)(285,-120){2} \Text(320,-40)[]{$\times$}
\DashArrowArcn(285,-120)(45,90,0){2}
\DashArrowArc(285,-120)(45,90,180){2}
\Text(350,-130)[]{$\nu_{aR}$} \Text(210,-130)[]{$ \nu^c_{bR} $}
\Text(260,-130)[]{$ l_{dL}$} \Text(305,-130)[]{$ l_{cR}$}
\Text(240,-130)[]{$\la^\prime$} \Text(285,-165)[]{$\times$}
\Text(295,-155)[]{$\rho_0^\prime$} \Text(285,-110)[]{$\ga$}
\Text(330,-130)[]{$\ga $}
\Text(333,-95)[]{$ \rho^{- \prime}_3$} 
\Text(244,-90)[]{$ \rho^+_1$} \DashArrowLine(250,-40)(285,-75){2}
\DashArrowLine(320,-40)(285,-75){2} \Text(320,-40)[]{$\times$}
\Text(250,-40)[]{$\times$} \Text(250,-60)[]{$\rho^{0}$}
\Text(320,-60)[]{$\rho^{0 ^\prime}$}
\Text(285,-85)[]{$ \propto g^2$}

\ArrowLine(410,-120)(440,-120) \ArrowLine(530,-120)(560,-120)
\ArrowLine(485,-120)(440,-120) \ArrowLine(530,-120)(485,-120)
\DashArrowLine (485,-120)(485,-165){2} \Text(520,-40)[]{$\times$}
\DashArrowArcn(485,-120)(45,90,0){2}
\DashArrowArc(485,-120)(45,90,180){2}
\Text(550,-130)[]{$\nu_{aR}$} \Text(410,-130)[]{$ \nu^c_{bR} $}
\Text(460,-130)[]{$ l^c_{dR}$} \Text(505,-130)[]{$ l^c_{cL}$}
\Text(440,-130)[]{$\ga$} \Text(485,-165)[]{$\times$}
\Text(495,-155)[]{$\rho_o^\prime$} \Text(485,-110)[]{$\ga$}
\Text(530,-130)[]{$\la^\prime $}
\Text(533,-95)[]{$ \rho^{-}_1$} 
\Text(444,-90)[]{$\rho^{\prime +}_3$}
\DashArrowLine(450,-40)(485,-75){2}
\DashArrowLine(520,-40)(485,-75){2} \Text(520,-40)[]{$\times$}
\Text(450,-40)[]{$\times$} \Text(450,-60)[]{$\rho^{0}$}
\Text(520,-60)[]{$\rho^{0 ^\prime}$}
\Text(485,-85)[]{$  \propto g^2$}
\end{picture}
\end{center}

\caption[]{\label{figh1} One-loop contribution  to the  mass
matrix $ M_R$}
\end{figure}

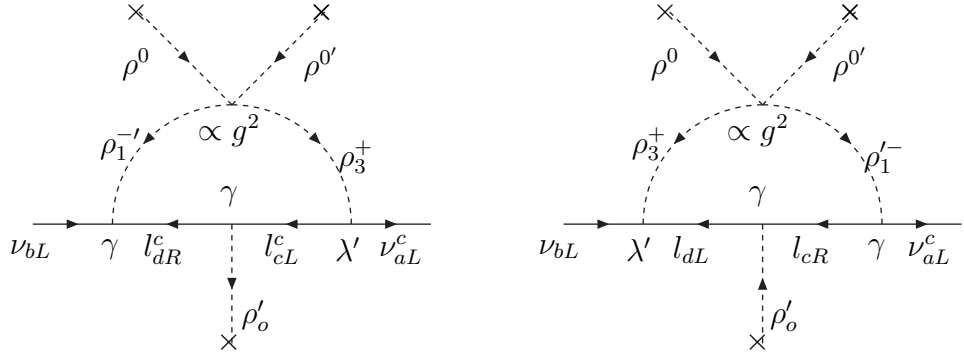
\begin{figure}[htbp]
\begin{center}
\begin{picture}(600,120)(150,-150)
\ArrowLine(210,-120)(240,-120) \ArrowLine(330,-120)(360,-120)
\ArrowLine(285,-120)(240,-120) \ArrowLine(330,-120)(285,-120)
\DashArrowLine (285,-120)(285,-165){2} \Text(320,-40)[]{$\times$}
\DashArrowArcn(285,-120)(45,90,0){2}
\DashArrowArc(285,-120)(45,90,180){2}
\Text(350,-130)[]{$\nu^c_{aL}$} \Text(210,-130)[]{$ \nu_{bL} $}
\Text(260,-130)[]{$ l^c_{dR}$} \Text(305,-130)[]{$ l^c_{cL}$}
\Text(240,-130)[]{$\ga$} \Text(285,-165)[]{$\times$}
\Text(295,-155)[]{$\rho_o^\prime$} \Text(285,-110)[]{$\ga$}
\Text(330,-130)[]{$\la^\prime $}
\Text(333,-95)[]{$ \rho^+_3$} 
\Text(244,-90)[]{$ \rho^{- \prime}_1$}
\DashArrowLine(250,-40)(285,-75){2}
\DashArrowLine(320,-40)(285,-75){2} \Text(320,-40)[]{$\times$}
\Text(250,-40)[]{$\times$} \Text(250,-60)[]{$\rho^{0}$}
\Text(320,-60)[]{$\rho^{0 ^\prime}$}
\Text(285,-85)[]{$\propto g^2$}

\ArrowLine(410,-120)(440,-120) \ArrowLine(530,-120)(560,-120)
\ArrowLine(485,-120)(440,-120) \ArrowLine(530,-120)(485,-120)
\DashArrowLine (485,-165)(485,-120){2} \Text(520,-40)[]{$\times$}
\DashArrowArcn(485,-120)(45,90,0){2}
\DashArrowArc(485,-120)(45,90,180){2}
\Text(550,-130)[]{$\nu^c_{aL}$} \Text(410,-130)[]{$ \nu_{bL} $}
\Text(460,-130)[]{$ l_{dL}$} \Text(505,-130)[]{$ l_{cR}$}
\Text(440,-130)[]{$\la^\prime$} \Text(485,-165)[]{$\times$}
\Text(495,-155)[]{$\rho_o^\prime$} \Text(485,-110)[]{$\ga$}
\Text(530,-130)[]{$\ga $}
\Text(533,-95)[]{$ \rho^{\prime -}_1$} 
\Text(444,-90)[]{$\rho^{+}_3$} \DashArrowLine(450,-40)(485,-75){2}
\DashArrowLine(520,-40)(485,-75){2} \Text(520,-40)[]{$\times$}
\Text(450,-40)[]{$\times$} \Text(450,-60)[]{$\rho^{0}$}
\Text(520,-60)[]{$\rho^{0 ^\prime}$} \Text(485,-85)[]{$  \propto
g^2$}
\end{picture}
\end{center}

\caption[]{\label{figh2} One-loop contribution  to the  mass
matrix $ M_L$}
\end{figure}

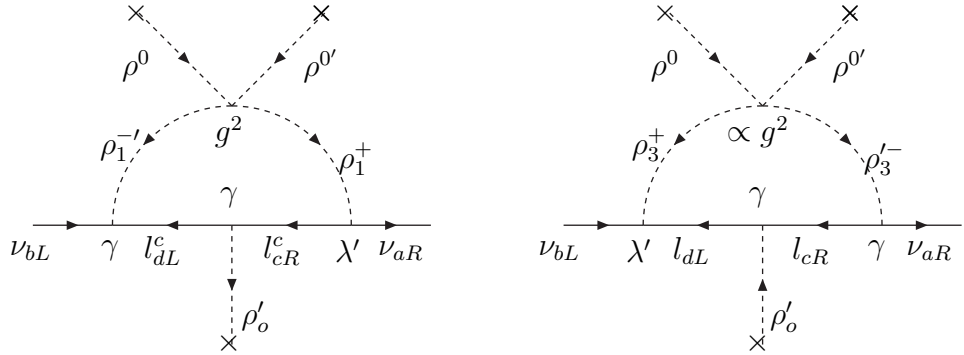
\begin{figure}[htbp]
\begin{center}
\begin{picture}(600,120)(150,-150)
\ArrowLine(210,-120)(240,-120) \ArrowLine(330,-120)(360,-120)
\ArrowLine(285,-120)(240,-120) \ArrowLine(330,-120)(285,-120)
\DashArrowLine (285,-120)(285,-165){2} \Text(320,-40)[]{$\times$}
\DashArrowArcn(285,-120)(45,90,0){2}
\DashArrowArc(285,-120)(45,90,180){2}
\Text(350,-130)[]{$\nu_{aR}$} \Text(210,-130)[]{$ \nu_{bL} $}
\Text(260,-130)[]{$ l^c_{dL}$} \Text(305,-130)[]{$ l^c_{cR}$}
\Text(240,-130)[]{$\ga$} \Text(285,-165)[]{$\times$}
\Text(295,-155)[]{$\rho_o^\prime$} \Text(285,-110)[]{$\ga$}
\Text(330,-130)[]{$\la^\prime $}
\Text(333,-95)[]{$ \rho^+_1$} 
\Text(244,-90)[]{$ \rho^{- \prime}_1$}
\DashArrowLine(250,-40)(285,-75){2}
\DashArrowLine(320,-40)(285,-75){2} \Text(320,-40)[]{$\times$}
\Text(250,-40)[]{$\times$} \Text(250,-60)[]{$\rho^{0}$}
\Text(320,-60)[]{$\rho^{0 ^\prime}$} \Text(285,-85)[]{$  g^2$}

\ArrowLine(410,-120)(440,-120) \ArrowLine(530,-120)(560,-120)
\ArrowLine(485,-120)(440,-120) \ArrowLine(530,-120)(485,-120)
\DashArrowLine (485,-165)(485,-120){2} \Text(520,-40)[]{$\times$}
\DashArrowArcn(485,-120)(45,90,0){2}
\DashArrowArc(485,-120)(45,90,180){2}
\Text(550,-130)[]{$\nu_{aR}$} \Text(410,-130)[]{$ \nu_{bL} $}
\Text(460,-130)[]{$ l_{dL}$} \Text(505,-130)[]{$ l_{cR}$}
\Text(440,-130)[]{$\la^\prime$} \Text(485,-165)[]{$\times$}
\Text(495,-155)[]{$\rho_o^\prime$} \Text(485,-110)[]{$\ga$}
\Text(530,-130)[]{$\ga $}
\Text(533,-95)[]{$ \rho^{\prime -}_3$} 
\Text(444,-90)[]{$\rho^{+}_3$} \DashArrowLine(450,-40)(485,-75){2}
\DashArrowLine(520,-40)(485,-75){2} \Text(520,-40)[]{$\times$}
\Text(450,-40)[]{$\times$} \Text(450,-60)[]{$\rho^{0}$}
\Text(520,-60)[]{$\rho^{0 ^\prime}$}
\Text(485,-85)[]{$  \propto g^2$}
\end{picture}
\end{center}
\caption[]{\label{figh3} One-loop contribution  to the  mass
matrix $ M_D$}
\end{figure}

With the couplings given in (\ref{t9181}), the right- and
left-handed neutrino mass matrices are given by a sum of two
one-loop diagrams, shown in Figs.\ref{figh1} and \ref{figh2},
respectively:
 \bea
i(M_L)_{ab}P_L&=&\int\fr{d^4p}{(2\pi)^4}\left(i2\la^\prime_{ac}P_L\right)
\fr{i(p\!\!\!/+m_c)}{p^2-m^2_c}\left(i\ga_{cd}\fr{v}{\sqrt{2}}P_R\right)
\fr{i(p\!\!\!/+m_d)}{p^2-m^2_d}\crn
&\times&(i\ga^{*}_{bd}P_L)\fr{-1}{(p^2-m^2_{\rho_1^{ \prime +}})
(p^2-m^2_{\rho^{ -}_3})}\left(ig^2v v^\prime\right)\crn
&+&\int\fr{d^4p}{(2\pi)^4}\left(i\ga^{*}_{ac}P_L\right)
\fr{i(-p\!\!\!/+m_c)}{p^2-m^2_c}\left(i\ga_{dc}\fr{v}{\sqrt{2}}P_R\right)
\fr{i(-p\!\!\!/+m_d)}{p^2-m^2_d}\crn
&\times&(i2\la^{\prime}_{bd}P_L)\fr{-1}{(p^2-m^2_{\rho^{\prime
-}_1}) (p^2-m^2_{\rho_3^+})}\left(ig^2v
v^\prime\right)\label{eq1}  \crn &=& i\sqrt{2}g^2 v
\la^\prime_{ab} P_L [m_b^2I(m_b^2,m_{\rho_1^{\prime
+}}^2,m_{\rho_3^-}^2)-m_a^2I(m_a^2,m_{\rho_1{^\prime
+}}^2,m_{\rho_3^-}^{2})] \crn \eea with $a,b$ are not summed.

Similarly, we have \be (M_R)_{ab}  = - (M_L)_{ab}. \ee
 Because of $m_{\rho_1{^\prime +}} =m_W ,m_{\rho_3^-}=0$, we obtain
 \bea
 I(m^2,m_{\rho_3}^2,m_{\rho_3}^2)&\simeq &-\fr{i}{16 \pi^2 m^2},\crn
 I(m_a^2,m_{\rho_1{^\prime
+}}^2,m_{\rho_3^-}^{2})&\simeq&
-\fr{i}{16\pi^2}\fr{1}{m^2_a-m^2_{\rho_1^{\prime
-}}}\left(1-\fr{m^2_{\rho_1^{\prime -}}
}{m^2_a-m^2_{\rho_1^{\prime -}}}\ln\fr{m^2_a}{m^2_{\rho_1^{\prime
-}}}\right), \label{hung} \\ \  & & a=  e, \mu, \tau \nn \eea
 With the functions
given in Eq.(\ref{hung}), the one loop correction to the mass
matrix $M_L$ can be written as \bea (M_L)_{ab}&\propto& -
(M_R)_{ab} \crn & = & \sqrt{2}\fr{g^2}{16 \pi^2} \la^\prime_{ab}v
\left[\fr{m_b^2}{m_{\rho_1^{\prime -}}^2}
\left(1-\ln\fr{m^2_b}{m^2_{\rho_1^{\prime -}}}\right)-\fr{m_a^2}{m_{\rho_1^{\prime -}}^2}
\left(1-\ln\fr{m^2_a}{m^2_{\rho_1^{\prime -}}}\right)\right] \crn
&\simeq& (M_{D}^{tree})_{ab} \propto v \crn \eea

Thus, the one-loop correction leads to the relationship $ M_L =-
M_R$, which is similar to the case of non-supersymmetric
economical 3-3-1 model \cite{dls1}. These mass matrices are
proportional to the  value $v$ but they are suppressed by an extra
factor $\frac{g^2}{16 \pi^2}$. Hence, the dominant matrix is
$M_D$, and it can be diagonalized by biunitary transformation as
the same as  in the non-supersymmetric economical 3-3-1 model
\cite{dls1}. This gives six different values: two lights and four
heavies. Let us consider the one-loop contribution to the Dirac
neutrino masses. Applying the Feynman rules to the
Fig.\ref{figh3}, we obtain   contribution to the mass matrix $M_D$
of the form
 \bea -i
(M^{rad}_{D})_{ab}P_L&=&\int\fr{d^4p}{(2\pi)^4}\left(-i2\la^\prime_{ac}P_L\right)
\fr{i(p\!\!\!/+m_c)}{p^2-m^2_c}\left(i\ga_{cd}\fr{v}{\sqrt{2}}P_R\right)
\fr{i(p\!\!\!/+m_d)}{p^2-m^2_d}\crn
&\times&(i\ga^{*}_{bd}P_L)\fr{-1}{(p^2-m^2_{\rho^+_1})(p^2-m^2_{\rho^{\prime
-}_1})} \left(g^2vv^\prime \right)\crn
&+&\int\fr{d^4p}{(2\pi)^4}\left(i\ga^{*}_{ac}P_L\right)
\fr{i(-p\!\!\!/+m_c)}{p^2-m^2_c}\left(i\ga_{dc}\fr{v}{\sqrt{2}}P_R\right)
\fr{i(-p\!\!\!/+m_d)}{p^2-m^2_d}\crn
&\times&(i2\la^\prime_{bd}P_L)\fr{-1}{(p^2-m^2_{\rho^+_2})(p^2-m^2_{\rho^{\prime
-}_2})} \left(g^2vv^\prime \right).\label{eq2}\eea We rewrite the
above result as \bea (M^{rad}_{D})_{ab} &=& \fr{g^2}{ 16 \pi^2}
\la^\prime_{ab}v \left[1- \fr{m_a^2}{m_{\rho^{\prime -}}^2}
\left(1-\ln\fr{m^2_a}{m^2_{\rho_1^{\prime -}}}\right)\right] \crn
 \propto v. \label{ham1}\eea
 It is very interesting that the scale for one-loop
correction to the Dirac masses is  proportional to the expectation
values v, the same as that of the tree level. However, unlike the
case of the tree level, the mass matrix given in  (\ref{ham1}) is
non-antisymmetric in $a$ and $b$. Hence, after including the
one-loop correction to the Dirac neutrino mass, all  three
eigenvalues of the Dirac mass matrix are non-zero. On the other
hand,  the left and right handed neutrino mass matrices are gained
at the one-loop correction. However, there is no larger hierarchy
between $M_{L}, M_R$ and $M_D$.  It is difficult to obtain the
seesaw mechanism  in this scenario. To solve this puzzle, as in
the non-supersymmetric economical 3-3-1  model, it is necessary to
introduce a new mass of the GUT scale ${\mathcal M} \simeq
10^{16}$ GeV \cite{dls1}.

Below we shall show that, in the model with an inflationary
scenario, the type I seesaw mechanism  can appear naturally.

\section{The seesaw mechanism   in
supersymmetric economical 3-3-1 model with an inflationary
scenario} \label{seesaw} 
\begin{figure}[htbp]
\begin{center}
\begin{picture}(650,430)(-45,-320)

\ArrowLine(10,20)(40,20) \ArrowLine(130,20)(160,20)
\ArrowLine(40,20)(85,20) \ArrowLine(85,20)(130,20)
\DashArrowArcn(85,20)(45,90,0){2}
\DashArrowArcn(85,20)(45,180,90){2}

\Text(160,10)[]{$\nu_{a R}$} \Text(10,10)[]{$ \nu_{b R}^c $}
\Text(65,10)[]{$ \widetilde{\phi} $} \Text(105,10)[]{$
\widetilde{\phi}$} \Text(40,10)[]{$\mu^\prime_{ ob}$}
\Text(85,20)[]{$\times$}
 \Text(130,10)[]{$\mu^\prime_{ oa} $}
\Text(133,45)[]{$ \chi_3^{o \prime}$}
 \Text(44,50)[]{$\chi_3^{o \prime}$}
\Text(85,55)[]{$ \fr{g^{\prime 2}}{27}$} \Text(85,-25)[]{(a)}
\DashArrowLine(85,65)(120,90){2} \Text(120,90)[]{$\times$}
\DashArrowLine(50,90)(85,65){2} \Text(50,90)[]{$\times$}
\Text(45,100)[]{$ \chi_3^{0 \prime}$} \Text(120,100)[]{$ \chi_3^{0
\prime}$}


\ArrowLine(210,20)(240,20) \ArrowLine(330,20)(360,20)
\ArrowLine(240,20)(285,20) \ArrowLine(285,20)(330,20)
\DashArrowArcn(285,20)(45,90,0){2}
\DashArrowArcn(285,20)(45,180,90){2} \Text(360,10)[]{$\nu_{aR}$}
\Text(210,10)[]{$ \nu_{bR}^c $} \Text(260,10)[]{$\widetilde{\phi}
$} \Text(305,10)[]{$\widetilde{\phi} $}
\Text(240,10)[]{$\mu^\prime_{ob}$} \Text(285,20)[]{$\times$}
 \Text(330,10)[]{$\mu_{0a}^{0 \prime} $}
\Text(333,45)[]{$ \chi_3^{0 \prime }$ } 
\Text(244,50)[]{$ \chi_3^{0 \prime }$}
\DashArrowLine(285,65)(320,90){2}
\DashArrowLine(250,90)(285,65){2} \Text(320,90)[]{$\times$}
\Text(250,90)[]{$\times$} \Text(245,100)[]{$\chi_3^0$}
\Text(320,100)[]{$\chi_3^0$}\Text(285,-25)[]{(b)}

\Text(285,55)[]{$ - \fr{g^{\prime 2}}{54}$}


\ArrowLine(10,-120)(40,-120) \ArrowLine(130,-120)(160,-120)
\ArrowLine(40,-120)(85,-120) \ArrowLine(85,-120)(130,-120)
\DashArrowArcn(85,-120)(45,90,0){2}
\DashArrowArcn(85,-120)(45,180,90){2}
\Text(160,-130)[]{$\nu_{aR}$} \Text(10,-130)[]{$ \nu_{bR}^c $}
\Text(60,-130)[]{$ \widetilde{\phi}$}
\Text(105,-130)[]{$\widetilde{\phi}$}
\Text(40,-130)[]{$\mu_{0b}^\prime$} \Text(85,-120)[]{$\times$}
 \Text(130,-130)[]{$\mu_{0a}^\prime $}
\Text(133,-95)[]{$ \chi_3^{0\prime}$} \Text(44,-90)[]{$
\chi_{3}^{0 \prime}$} \Text(85,-85)[]{$  -\fr{g^{\prime 2}}{54}$}
\Text(85,-165)[]{(c)} \DashArrowLine(120,-50)(85,-75){2}
\Text(120,-50)[]{$\times$} \DashArrowLine(85,-75)(50,-50){2}
\Text(50,-50)[]{$\times$} \Text(45,-40)[]{$\chi_1^0 $}
\Text(115,-40)[]{$\chi_1^0$}


\ArrowLine(210,-120)(240,-120) \ArrowLine(330,-120)(360,-120)
\ArrowLine(240,-120)(285,-120) \ArrowLine(285,-120)(330,-120)
\DashArrowArcn(285,-120)(45,90,0){2}
\DashArrowArcn(285,-120)(45,180,90){2}
\Text(360,-130)[]{$\nu_{aR}$} \Text(210,-130)[]{$ \nu_{bR}^c $}
\Text(260,-130)[]{$\widetilde{\phi}$} \Text(305,-130)[]{$
\widetilde{\phi} $} \Text(240,-130)[]{$\mu_{0b}^\prime$}
\Text(285,-120)[]{$\times$}
 \Text(330,-130)[]{$\mu_{0a}^\prime $}
\Text(333,-95)[]{$ \chi_3^{0 \prime}$} 
\Text(244,-90)[]{$ \chi_3^{o \prime }$}
\DashArrowLine(250,-50)(285,-75){2}
\DashArrowLine(285,-75)(320,-50){2} \Text(320,-50)[]{$\times$}
\Text(250,-50)[]{$\times$} \Text(250,-40)[]{$\chi_1^{0 \prime}$}
\Text(320,-40)[]{$\chi_1^{0 \prime}$}\Text(285,-165)[]{(d)}
 \Text(285,-85)[]{$  \fr{g^{\prime 2}}{54}$}


\ArrowLine(10,-260)(40,-260) \ArrowLine(130,-260)(160,-260)
\ArrowLine(40,-260)(85,-260) \ArrowLine(85,-260)(130,-260)
\DashArrowArcn(85,-260)(45,90,0){2}
\DashArrowArcn(85,-260)(45,180,90){2}
\Text(160,-270)[]{$\nu_{aR}$} \Text(10,-270)[]{$ \nu_{b R}^c $}
\Text(65,-270)[]{$\widetilde{ \phi} $}
\Text(105,-270)[]{$\widetilde{ \phi}$} \Text(40,-270)[]{$\mu_{
0b}^\prime$} \Text(85,-260)[]{$\times$}
 \Text(130,-270)[]{$\mu_{0a}^\prime $}
\Text(133,-235)[]{$ \chi^{0 \prime}_3$}
 \Text(44,-230)[]{$ \chi_3^{0 \prime}$}
\Text(85,-225)[]{$   \fr{g^{\prime 2}}{27}$} \Text(85,-305)[]{(e)}
\DashArrowLine(85,-215)(120,-190){2} \Text(120,-190)[]{$\times$}
\DashArrowLine(50,-190)(85,-215){2} \Text(50,-190)[]{$\times$}
\Text(45,-180)[]{$\rho_2^0$} \Text(120,-180)[]{$\rho_2^{0}$}


\ArrowLine(210,-260)(240,-260) \ArrowLine(330,-260)(360,-260)
\ArrowLine(240,-260)(285,-260) \ArrowLine(285,-260)(330,-260)
\DashArrowArcn(285,-260)(45,90,0){2}
\DashArrowArcn(285,-260)(45,180,90){2} \Text(360,-270)[]{$
\nu_{aR}$} \Text(210,-270)[]{$ \nu_{bR}^c $} \Text(260,-270)[]{$
\widetilde{\phi} $} \Text(305,-270)[]{$ \widetilde{\phi }$}
\Text(240,-270)[]{$\mu_{0b}^\prime$} \Text(285,-260)[]{$\times$}
 \Text(330,-270)[]{$\mu_{0a}^\prime $}
\Text(333,-235)[]{$ \chi_3^{0 \prime}$} 
\Text(244,-230)[]{$ \chi_3^{0 \prime}$}
\DashArrowLine(320,-190)(285,-215){2}
\DashArrowLine(285,-215)(250,-190){2} \Text(320,-190)[]{$\times$}
\Text(250,-190)[]{$\times$} \Text(245,-180)[]{$\rho_2^{\prime o}$}
\Text(320,-180)[]{$\rho_2^{\prime o }$}\Text(285,-305)[]{(f)}

\Text(285,-225)[]{$   -\fr{g^{\prime 2}}{27}$}
\end{picture}
\end{center}
\caption[]{\label{early1} One-loop contribution  to the neutrino
mass matrix $M_R$ }
\end{figure}
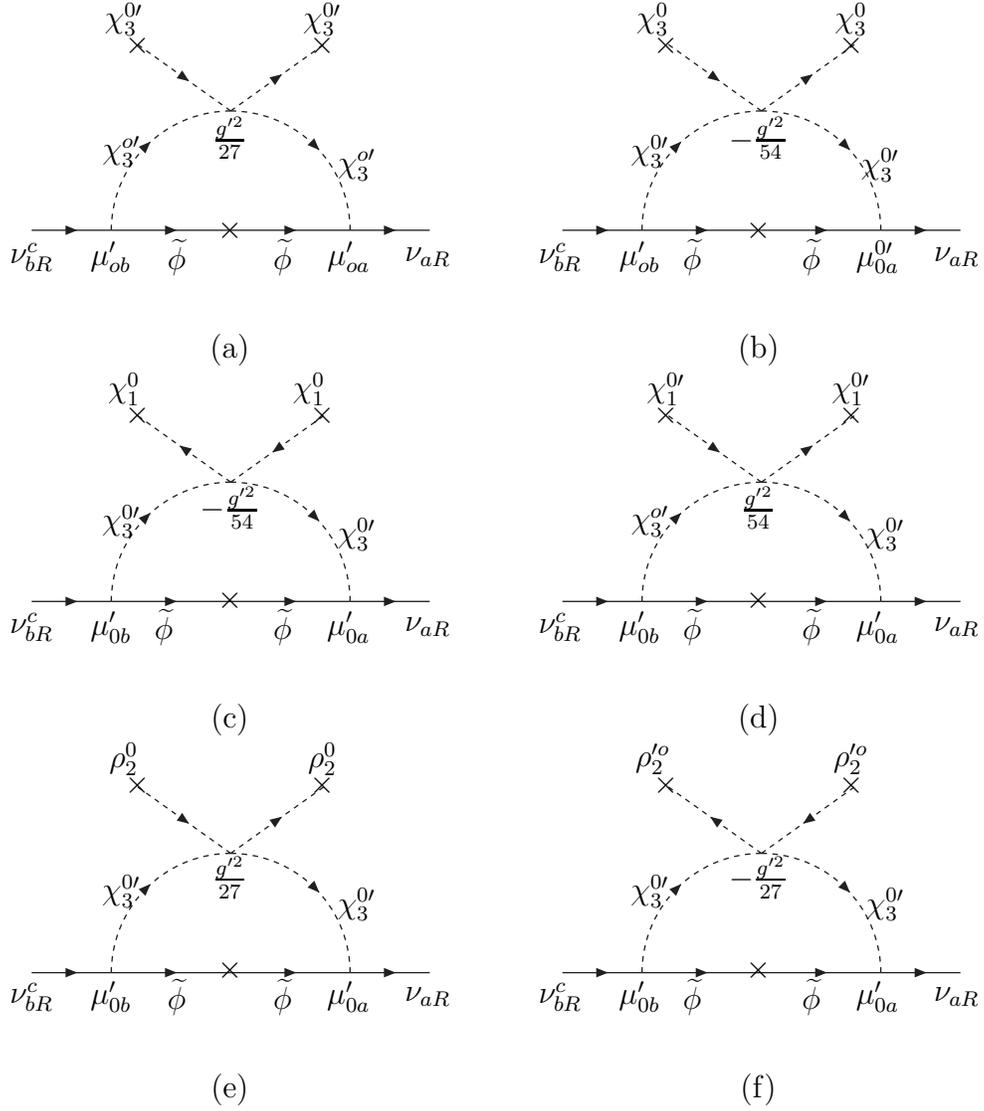

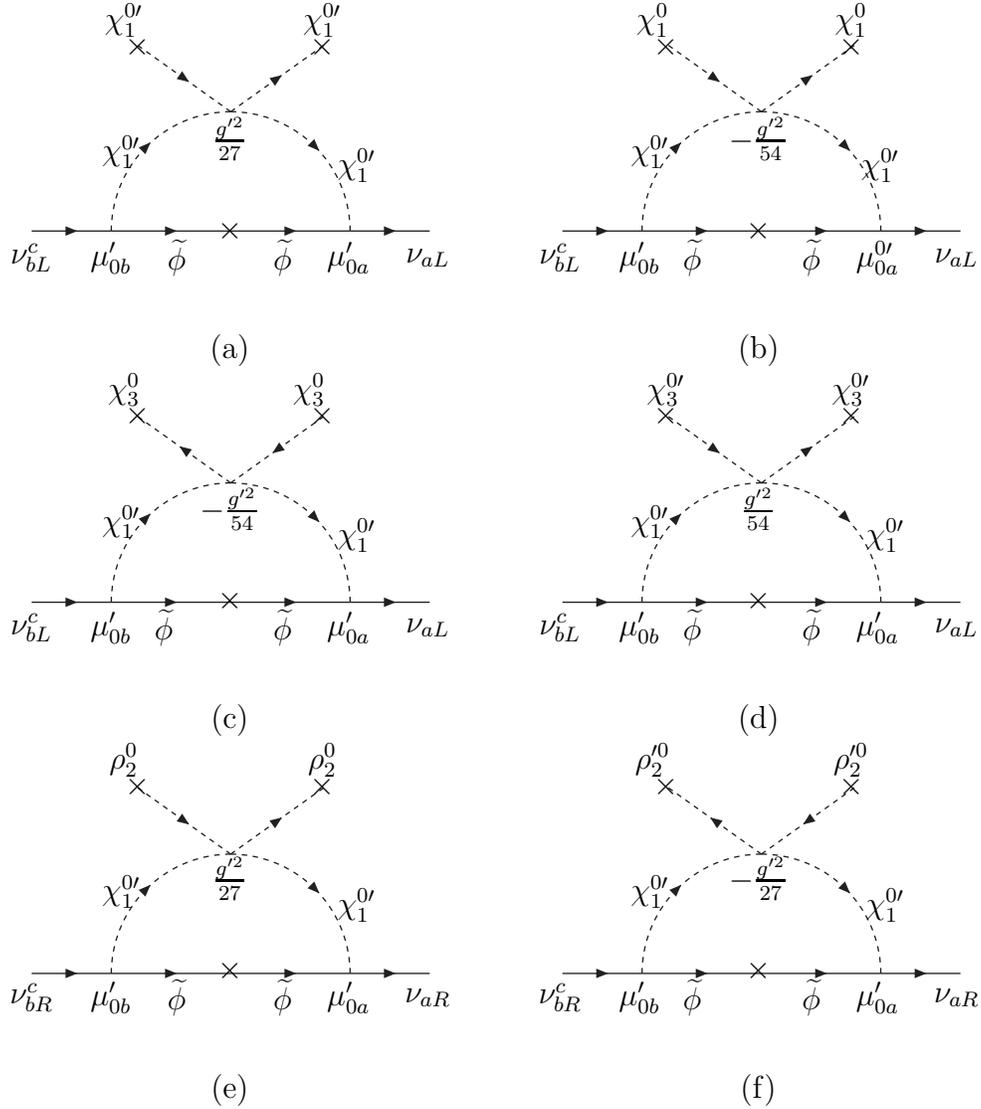
\begin{figure}[htbp]
\begin{center}
\begin{picture}(650,400)(-45,-320)

\ArrowLine(10,20)(40,20) \ArrowLine(130,20)(160,20)
\ArrowLine(40,20)(85,20) \ArrowLine(85,20)(130,20)
\DashArrowArcn(85,20)(45,90,0){2}
\DashArrowArcn(85,20)(45,180,90){2}

\Text(160,10)[]{$\nu_{a L}$} \Text(10,10)[]{$ \nu_{b L}^c $}
\Text(65,10)[]{$ \widetilde{\phi} $} \Text(105,10)[]{$
\widetilde{\phi}$} \Text(40,10)[]{$\mu^\prime_{ 0b}$}
\Text(85,20)[]{$\times$}
 \Text(130,10)[]{$\mu^\prime_{ 0a} $}
\Text(133,45)[]{$ \chi_1^{0 \prime}$}
 \Text(44,50)[]{$\chi_1^{0 \prime}$}
\Text(85,55)[]{$ \fr{g^{\prime 2}}{27}$} \Text(85,-25)[]{(a)}
\DashArrowLine(85,65)(120,90){2} \Text(120,90)[]{$\times$}
\DashArrowLine(50,90)(85,65){2} \Text(50,90)[]{$\times$}
\Text(45,100)[]{$ \chi_1^{0 \prime}$} \Text(120,100)[]{$ \chi_1^{0
\prime}$}


\ArrowLine(210,20)(240,20) \ArrowLine(330,20)(360,20)
\ArrowLine(240,20)(285,20) \ArrowLine(285,20)(330,20)
\DashArrowArcn(285,20)(45,90,0){2}
\DashArrowArcn(285,20)(45,180,90){2} \Text(360,10)[]{$\nu_{aL}$}
\Text(210,10)[]{$ \nu_{bL}^c $} \Text(260,10)[]{$\widetilde{\phi}
$} \Text(305,10)[]{$\widetilde{\phi} $}
\Text(240,10)[]{$\mu^\prime_{0b}$} \Text(285,20)[]{$\times$}
 \Text(330,10)[]{$\mu_{0a}^{0 \prime} $}
\Text(333,45)[]{$ \chi_1^{0 \prime }$ } \Text(244,50)[]{$
\chi_1^{0 \prime }$} \DashArrowLine(285,65)(320,90){2}
\DashArrowLine(250,90)(285,65){2} \Text(320,90)[]{$\times$}
\Text(250,90)[]{$\times$} \Text(245,100)[]{$\chi_1^0$}
\Text(320,100)[]{$\chi_1^0$}\Text(285,-25)[]{(b)}

\Text(285,55)[]{$ - \fr{g^{\prime 2}}{54}$}


\ArrowLine(10,-120)(40,-120) \ArrowLine(130,-120)(160,-120)
\ArrowLine(40,-120)(85,-120) \ArrowLine(85,-120)(130,-120)
\DashArrowArcn(85,-120)(45,90,0){2}
\DashArrowArcn(85,-120)(45,180,90){2}
\Text(160,-130)[]{$\nu_{aL}$} \Text(10,-130)[]{$ \nu_{bL}^c $}
\Text(60,-130)[]{$ \widetilde{\phi}$}
\Text(105,-130)[]{$\widetilde{\phi}$}
\Text(40,-130)[]{$\mu_{0b}^\prime$} \Text(85,-120)[]{$\times$}
 \Text(130,-130)[]{$\mu_{0a}^\prime $}
\Text(133,-95)[]{$ \chi_1^{0\prime}$} \Text(44,-90)[]{$
\chi_{1}^{0 \prime}$} \Text(85,-85)[]{$  -\fr{g^{\prime 2}}{54}$}
\Text(85,-165)[]{(c)} \DashArrowLine(120,-50)(85,-75){2}
\Text(120,-50)[]{$\times$} \DashArrowLine(85,-75)(50,-50){2}
\Text(50,-50)[]{$\times$} \Text(45,-40)[]{$\chi_3^0$}
\Text(115,-40)[]{$\chi_3^0$}


\ArrowLine(210,-120)(240,-120) \ArrowLine(330,-120)(360,-120)
\ArrowLine(240,-120)(285,-120) \ArrowLine(285,-120)(330,-120)
\DashArrowArcn(285,-120)(45,90,0){2}
\DashArrowArcn(285,-120)(45,180,90){2}
\Text(360,-130)[]{$\nu_{aL}$} \Text(210,-130)[]{$ \nu_{bL}^c $}
\Text(260,-130)[]{$\widetilde{\phi}$} \Text(305,-130)[]{$
\widetilde{\phi} $} \Text(240,-130)[]{$\mu_{0b}^\prime$}
\Text(285,-120)[]{$\times$}
 \Text(330,-130)[]{$\mu_{0a}^\prime $}
\Text(333,-95)[]{$ \chi_1^{0 \prime}$} \Text(244,-90)[]{$
\chi_1^{0 \prime }$} \DashArrowLine(250,-50)(285,-75){2}
\DashArrowLine(285,-75)(320,-50){2} \Text(320,-50)[]{$\times$}
\Text(250,-50)[]{$\times$} \Text(250,-40)[]{$\chi_3^{0 \prime}$}
\Text(320,-40)[]{$\chi_3^{0 \prime}$}\Text(285,-165)[]{(d)}
 \Text(285,-85)[]{$  \fr{g^{\prime 2}}{54}$}


\ArrowLine(10,-260)(40,-260) \ArrowLine(130,-260)(160,-260)
\ArrowLine(40,-260)(85,-260) \ArrowLine(85,-260)(130,-260)
\DashArrowArcn(85,-260)(45,90,0){2}
\DashArrowArcn(85,-260)(45,180,90){2}
\Text(160,-270)[]{$\nu_{aR}$} \Text(10,-270)[]{$ \nu_{b R}^c $}
\Text(65,-270)[]{$\widetilde{ \phi} $}
\Text(105,-270)[]{$\widetilde{ \phi}$} \Text(40,-270)[]{$\mu_{
0b}^\prime$} \Text(85,-260)[]{$\times$}
 \Text(130,-270)[]{$\mu_{0a}^\prime $}
\Text(133,-235)[]{$ \chi^{0 \prime}_1$}
 \Text(44,-230)[]{$ \chi_1^{0 \prime}$}
\Text(85,-225)[]{$   \fr{g^{\prime 2}}{27}$} \Text(85,-305)[]{(e)}
\DashArrowLine(85,-215)(120,-190){2} \Text(120,-190)[]{$\times$}
\DashArrowLine(50,-190)(85,-215){2} \Text(50,-190)[]{$\times$}
\Text(45,-180)[]{$\rho_2^0$} \Text(120,-180)[]{$\rho_2^{0}$}


\ArrowLine(210,-260)(240,-260) \ArrowLine(330,-260)(360,-260)
\ArrowLine(240,-260)(285,-260) \ArrowLine(285,-260)(330,-260)
\DashArrowArcn(285,-260)(45,90,0){2}
\DashArrowArcn(285,-260)(45,180,90){2} \Text(360,-270)[]{$
\nu_{aR}$} \Text(210,-270)[]{$ \nu_{bR}^c $} \Text(260,-270)[]{$
\widetilde{\phi} $} \Text(305,-270)[]{$ \widetilde{\phi }$}
\Text(240,-270)[]{$\mu_{0b}^\prime$} \Text(285,-260)[]{$\times$}
 \Text(330,-270)[]{$\mu_{0a}^\prime $}
\Text(333,-235)[]{$ \chi_1^{0 \prime}$} \Text(244,-230)[]{$
\chi_1^{0 \prime}$} \DashArrowLine(320,-190)(285,-215){2}
\DashArrowLine(285,-215)(250,-190){2} \Text(320,-190)[]{$\times$}
\Text(250,-190)[]{$\times$} \Text(245,-180)[]{$\rho_2^{\prime 0}$}
\Text(320,-180)[]{$\rho_2^{\prime 0 }$}\Text(285,-305)[]{(f)}

\Text(285,-225)[]{$   -\fr{g^{\prime 2}}{27}$}
\end{picture}
\end{center}
\caption[]{\label{early2} One-loop contribution  to the neutrino
mass matrix $M_L$ }
\end{figure}

We have constructed a hybrid inflationary
scheme based on a realistic supersymmetric 3-3-1 model by adding a
singlet superfield $\widehat{\Phi}$ which plays the role of the
inflation, namely the inflaton superfield
 \cite{LHIn}. Let us recall that the inflationary potential is given by
 \be
W_{inf}(\widehat{\Phi},\widehat{\chi},\widehat{\chi}^\prime)=\alpha
\widehat{\Phi} \widehat{\chi }\widehat{\chi}^\prime -\mu^2
\widehat{\Phi}.\label{Poten}\ee The superpotential related to the
neutrino masses is \be W_{neut}= \mu_{0a}^\prime
\widehat{L}_a\widehat{\chi}^\prime \widehat{\phi} \ee Integrating
out the superspace gives the relevant interaction Lagrangian for
the one-loop correction to  neutrino mass  \bea L_{int}& =&
\mu_{0a}^\prime \nu_{aL} \widetilde{\phi} \chi_1^{'0} +
\mu_{0a}^\prime \nu_{aR}^c  \widetilde{\phi} \chi_3^{'0} + H.c.,
\label{th2} \\ V_{Higgs}^{rel.} &=& \al^2 (\chi \chi^\prime)^2
\label{inf1}\eea Besides the relevant Higgs self-coupling given in
Eq.(\ref{inf1}), there is  another Higgs potential contributing to
the neutrino mass at the one-loop correction, namely \bea
V_D&=&\frac{g^{\prime2}}{12} \left(-\frac{1}{3} \chi^{\dagger}
\chi + \frac{1}{3} \chi^{\prime \dagger} \chi^{\prime}+
\frac{2}{3} \rho^{\dagger}\rho - \frac{2}{3} \rho^{\prime
\dagger}\rho^{\prime} \right)^2\crn &
+&\frac{g^2}{8}(\chi^\dagger_i\lambda^b_{ij}\chi_j-
\chi^{\prime\dagger}_i\lambda^{*b}_{ij}\chi^\prime_j+
\rho^\dagger_i\lambda^b_{ij}\rho_j-
\rho^{\prime\dagger}_i\lambda^{*b}_{ij}\rho^\prime_j)^2
\label{inf2}\eea with $g^\prime, g$ are the gauge couplings of
$U(1), SU(3)_L$ groups, respectively. Because of this, the
$g^\prime$ coupling constant is the co-variant function of energy
and the $g$ coupling constant is the contravariant function  of
energy. At the inflationary  and preheating times, the $g^\prime $
coupling constant is dominated  and  we will ignore the self-Higgs
coupling in the second line of  Eq.(\ref{inf2}). On the other
hand, requiring  the nonadiabatic string contribution to the
quadrupole to be less than 10\%, the coupling $\al$ belongs to
$10^{-4} \div 10^{-8} $ \cite{LHIn}. If we compare this value with
that of $g^\prime$ coupling constant at the early time of the
universe, the values of $\al$ coupling is tiny enough to ignore
the Higgs self-coupling given in Eq.(\ref{inf1}).  In short, at
the inflationary  and preheating times, the Lagrangian related to
the one-loop correction to neutrino mass is given by\bea L_{int}&
=& \mu_{0a}^\prime \nu_{aL} \widetilde{\phi} \chi_1^{'0} +
\mu_{0a}^\prime \nu_{aR}^c
\widetilde{\phi} \chi_3^{'0} + H.c., \label{th1} \\
V_D^{U(1)}&=&\frac{g^{\prime2}}{12} \left(-\frac{1}{3}
\chi^{\dagger} \chi + \frac{1}{3} \chi^{\prime \dagger}
\chi^{\prime}+ \frac{2}{3} \rho^{\dagger}\rho - \frac{2}{3}
\rho^{\prime \dagger}\rho^{\prime} \right)^2 \eea
 At  the one-loop order, there is no  correction to the mass matrix  $M_D$
  but  there is   correction to the mass matrices $M_L$ and $M_R$  given in Figs.
 \ref{early1} and \ref{early2}

\vspace*{1cm}

We assume that the vacuum expectation values $w, u, v$ are the
same as $w^\prime, u^\prime, v^\prime$, respectively.  With this
assumption, the contributions from diagrams \ref{early2} (c) and
(d) are canceled by each other and similarly for diagrams
\ref{early2} (e) and (f).
 Hence, the total contribution to the neutrino mass matrix $M_L$ is obtained
 from   diagrams \ref{early2} (a) and (b) as follows:
 [see  (\ref{ketqua})]
\bea -i
M_{Lab}^{inf}P_L&=&\int\fr{d^4p}{(2\pi)^4}\left(i\mu_{oa}^\prime
P_L\right)
\fr{i(p\!\!\!/+m_{\widetilde{\phi}})}{p^2-m^2_{\widetilde{\phi}}}(-im_{\widetilde{\phi}})
\fr{i(p\!\!\!/+m_{\widetilde{\phi}})}{p^2-m^2_{\widetilde{\phi}}}\crn
&\times&(i\mu_{0a}^{\prime*}P_L)\fr{-1}{(p^2-m^2_{\chi_1^\prime})^2
}\left(i u^{ 2}\fr{g^{\prime 2}}{54}\right  ) \crn
&=&2m_{\widetilde{\phi}} \fr{g^{\prime 2}}{54}\mu^{\prime
\ast}_{0a}\mu_{0b}^\prime u^{ 2} P_L \int \fr{d^4
p}{(2\pi)^4}\fr{p^2}{(p^2-m_{\chi_1^\prime}^2)^2(p^2-m_{\widetilde{\phi}}^2)^2}
 \crn &+&2m_{\widetilde{\phi}} \fr{g^{\prime 2}}{54}\mu^{\prime
\ast}_{0a}\mu_{0b}^\prime u^{ 2} P_L\int \fr{d^4
p}{(2\pi)^4}\fr{m_{\widetilde{\phi}}^2}{(p^2-m_{\chi_1^\prime}^2)^2(p^2-m_{\widetilde{\phi}}^2)^2}
 \crn &=&m_{\widetilde{\phi}}\fr{g^{\prime 2}}{27}\mu^{\prime
\ast}_{0a}\mu_{0b}^\prime u^{ 2} P_L\left[I( m_{
\widetilde{\phi}}^2,m_{\chi^\prime}^2)+m_{\widetilde{\phi}}^2 I_1(
m_{\widetilde{\phi}}^2,m_{\chi^\prime}^2)\right] \label{inft1}\eea
Note that $\widetilde{\phi}$ is a super partner of inflaton; hence
their mass must be larger than those of inflaton. It means that
$m_{\widetilde{\phi}}\gg m_{\chi_1}^\prime$. If we take that the
ratio of $m_{\chi_1}^\prime$ to $m_{\widetilde{\phi}}$ is of the
order $10^{-x}$, we obtain \bea I(
m_{\widetilde{\phi}}^2,m_{\chi^\prime}^2)& \simeq& - \fr{i}{16
\pi^2 m^2_{m_{\widetilde{\phi}}^2}}, \crn I_1(
m_{\widetilde{\phi}}^2,m_{\chi^\prime}^2) &\simeq& - \fr{i}{16
\pi^2 m_{\widetilde{\phi}}^4}\left( 2- x \ln 10\right)
\label{inft2}\eea Substitution of Eq. (\ref{inft2}) into
Eq.(\ref{inft1}) gives \be M_{Lab}^{inf} \simeq  - \fr{i}{16 \pi^2
}\fr{g^{\prime 2}}{27}\mu^{\prime \ast}_{0a}\mu_{0b}^\prime
\fr{u^2}{m_{\widetilde{\phi}} } \ee Making similar steps to the
mass matrix $M_R$, we obtain the result \be M_{Rab}^{inf} \simeq -
\fr{i}{16 \pi^2 }\fr{g^{\prime 2}}{27}\mu^{\prime
\ast}_{0a}\mu_{0b}^\prime \fr{w^2}{m_{\widetilde{\phi}}} \ee The
neutrino masses are the eigenvalues of the matrix \be \left(
  \begin{array}{cc}
   M_{Lab}^{inf} & M_D ^T\\
    M_D& M_{Rab}^{inf} \\
  \end{array}
\right) \ee Because of the condition $w^\prime, w  \gg u^\prime,
u, v^\prime, v$ and $u^\prime, u \ll v^\prime, v$ [see
Eq.(\ref{coduu})] and
 ($ M_{R} \propto w^2, M_{D} \propto v^2, M_{L} \propto
 u^2$),
 we obtain a hierarchy in  values of the elements of  the neutrino mass:
  \be M_{Rab}^{inf}\gg M_D \gg M_{Lab}^{inf} \ee The heavy and
light eigenvectors are found to be  diagonalize  the matrices: \be
m_{R} = M_{Rab}^{inf}, \hs  m_\nu = M_D M_{Rab}^{inf -1}M_D^T. \ee
Let us mention again that in the framework of the
non-supersymmetric economical models as well as the supersymmetric
version without inflationary scenario, to get  successful neutrino
masses, it is necessary to introduce a new mass of the GUT scale
${\mathcal M} \simeq 10^{16}$ GeV \cite{dls1}. While in the
supersymmetric  model with an inflationary scenario, with the help
of the interactions among the inflaton and right handed neutrinos
(\ref{th2}), the above puzzle is solved. Thus the inflaton with
mass around $10^{17}$ GeV plays the role of new physics in the
economical models  with the inflationary scenario.

\section{Non-thermal leptogenesis via inflaton decay}
\label{nontermal}
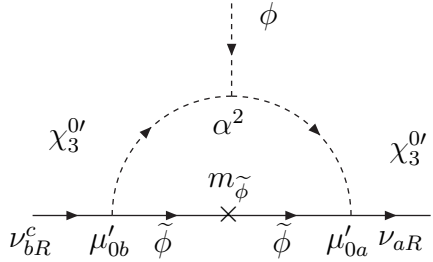
\begin{figure}[htbp]
\begin{center}
\begin{picture}(650,120)(40,-150)
\ArrowLine(210,-120)(240,-120) \ArrowLine(330,-120)(360,-120)
\ArrowLine(240,-120)(285,-120) \ArrowLine(285,-120)(330,-120)
\DashArrowArcn(285,-120)(45,90,0){2}
\DashArrowArcn(285,-120)(45,180,90){2}
\Text(350,-130)[]{$\nu_{aR}$} \Text(210,-130)[]{$ \nu^c_{bR} $}
\Text(260,-130)[]{$ \widetilde{\phi}$} \Text(305,-130)[]{$
\widetilde{\phi}$} \Text(240,-130)[]{$\mu_{0b}^\prime$}
\Text(285,-120)[]{$\times$}
\Text(285,-110)[]{$m_{\widetilde{\phi}}$}
\Text(330,-130)[]{$\mu_{0a}^\prime $} \Text(353,-95)[]{$
\chi_3^{0\prime}$} \Text(224,-90)[]{$ \chi_3^{0\prime}$}
 \DashArrowLine(285,-40)(285,-75){2}  \Text(300,-45)[]{$\phi$}
\Text(285,-85)[]{$ \al^2 $}
\end{picture}
\end{center}
\caption[]{\label{inter} Feynman diagram for the process $\phi
\rightarrow \nu_R \nu_R$  }
\end{figure}

 \hs Let us  consider the non-thermal
leptongenesis scenario in our model. In the non-thermal
leptongenesis scenario, the right handed neutrinos are produced
through the direct non-thermal decay of the inflaton. In our
scenario, there is no interaction term which describes that decay
process at the tree level. However, the necessary interaction
arises  at the one-loop level. The relevant self-Higgs and
inflaton couplings is given by \be L_{thermal}=\left|\fr{\partial
W_{inf}}{\partial \chi}\right|^2+ \left|\fr{\partial
W_{inf}}{\partial \chi^\prime}\right|^2 = \al^2 \left( |\chi |^2
+|\chi ^{\prime}|^2 \right)\phi \label{ther}\ee From the
Lagrangian given in  (\ref{th1}) and (\ref{ther}), the effective
interaction relevant for the right handed neutrinos and inflaton
at the one-loop correction is given in Fig. \ref{inter}.

The effective Lagrangian for the process $\phi \rightarrow \nu_R
\nu_R$ is given by \be L_{\nu_R \nu_R \phi}= A_{eff}\phi \nu_R
\nu_R+H.c \ee where $A_{eff}$ stands for effective coupling, which
is obtained as\bea A_{eff} &\propto&2 m_{\widetilde{\phi}}\al^2
\mu^{\prime \ast}_{0a}\mu_{0b}^\prime [I(
m_{\widetilde{\phi}}^2,m_{\chi^\prime_3}^2)+m_{\widetilde{\phi}}^2I_1(
m_{\widetilde{\phi}}^2,m_{\chi^\prime_3}^2)]  \crn &\propto& 54
\fr{M_R }{g^{\prime 2} w^2} \al^2 \label{Aeff}\eea The inflaton
decay rate  is given by \be \Gamma (\phi \rightarrow \nu_R \nu_R)
\simeq \fr{ |A_{eff}|^2}{4 \pi} m_\phi \label{Gam}\ee with
$m_\phi$  the inflaton mass. The produced reheating temperature is
obtained by \be T_R = \left( \fr{45}{2 \pi^2
g_*}\right)^{\fr{1}{4}}(\Gamma M_P)^{\fr{1}{2}} \label{Tr}\ee
where $g^*$ is the effective degree of the freedom in the universe
at $T\sim M_R$. In our model, the effective degree of the freedom
is taken approximately 140 (for more details, see \cite{effg}).

\begin{figure}[htbp]
\includegraphics[width=6cm,height=6cm]{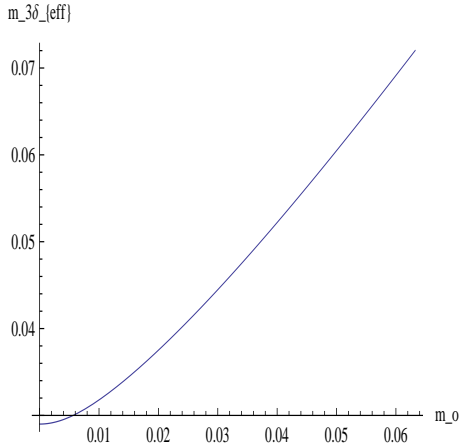}
\caption{\label{resubh1m4}The product of $m_{\nu 3} [\textrm{eV}]$
and $\delta_{eff} $ is a function of $ m_0$ by taking the maximal
CP violating phases ($\sin \delta_{13}=\sin \delta_{12}=1$).
\label{specsugrah}}
\end{figure}

We assume that the inflaton $\phi$ decays dominantly into a pair
of the lightest heavy Majorana neutrino, $\phi \rightarrow \nu_{R
1},\nu_{R1}$, and other decay modes including these into pair
$\nu_{R2}, \nu_{R3}$ are forbidden. The inflaton decays to
lightest heavy neutrino and that neutrino decay to charged leptons
and  Higgs reheats the Universe, producing not only the
lepton-number asymmetry but also entropy for thermal bath. The
interference between the tree-level decay amplitude and the
absorptive part of the one-loop diagram can lead to a lepton
asymmetry of the right order of magnitude to explain the observed
baryon asymmetry. The $N_1$ decays immediately after beying
produced by the inflaton decays and hence we obtain
lepton-to-entropy ratio \cite{nonther} \bea \fr{n_L}{s} \simeq \fr
3 2 \epsilon \times B_r \times \fr{T_R}{m_\phi}\label{th3} \eea
where $B_r$ is the branching ratio of the inflaton decay  into the
$N_1$ channel.
 The lepton asymmetry (in (Eq.\ref{th3})) is converted to the baryon
asymmetry through the ``sphaleron''  effects which is given by \be
\fr{n_B}{s} = a \fr{n_L}{s} \label{th4} \ee
 with $a = - \fr{8}{23}$ in the MSSM.
 The ratio of the lepton number to entropy density after preheating is
estimated to be \cite{nonther} \be \fr{n_B}{s}=-0.35 \times
\fr{3}{2}B_r(\phi \rightarrow  \nu_{R
1},\nu_{R1})\fr{T_R}{M_\phi}\times \epsilon. \ee
 The lepton asymmetry parameter $\epsilon$ is produced by the interference
  between the tree level and one-loop level
 of the $\nu_R \rightarrow l_L \rho $ or $\nu_R \rightarrow l_L \rho^\prime $ decay process.
  The   thermal leptogenesis scenario, in detail, in the
 economical 3-3-1 model  will be presented
 elsewhere \cite{Pre}. The CP violating parameter \cite{LCovi} is
 given by
 \be
 \epsilon =\fr{1}{(8\pi \la^\prime \la^{\prime \dagger})_{11}}\sum_{j=2,3}
 Im\left[( \la^\prime \la^{\prime \dagger})_{1j}^2\right]
 \left[ f(M^2_{Rj}/M^2_{R1})+2g(M_{Rj}^2/M_{R1}^2)\right]
 \ee
 with $f(x)$ and $g(x)$  the vertex and the wave functions, respectively.
 In the limit $x \gg 1$, the CP violating parameter $\epsilon$
 can be written as
 \be
 \epsilon = -\fr{3}{16 \pi (\la^{\prime}\la^{\prime \dagger})_{11}} \sum_{j=2,3}
 Im\left[( \la^\prime \la^{\prime \dagger})_{1j}^2\right]
  \fr{M_{R1}}{M_{Rj}}
 \ee
 As mentioned in the last section, we have  type I seesaw mechanism $
 m_\nu = M_D M_{R}^{-1}M_D^T = \la^\prime  M_{R}^{-1}\la^{\prime T} \langle\rho \rangle^2
 $,  hence the CP violating parameter can be written as
 \bea
 \epsilon &=-& \fr{3}{16 \pi}\fr{M_{R1}}{\langle\rho\rangle^2}\fr{Im [\la^\prime M_\nu^*
 \la^{\prime T}]}{(\la^{\prime}\la^{\prime \dagger})_{11}} \crn
 &=-&\fr{3}{16 \pi}\fr{m_{\nu 3}M_{R1}\delta_{eff}}{\langle\rho\rangle^2}
 \eea
 where the effective CP-violating phase $\delta_{eff}$ is given by
 \be
 \delta_{eff}= \fr {Im\left[\la_{13}^{\prime 2}+\fr{m_{\nu 2}}{m_{\nu 3} }\la_{12}^{\prime 2}
 +\fr{m_{ \nu 1}}{m_{\nu 3}}\la_{11}^{\prime 2}\right]}{|\la^{\prime}_{13}|^2
 +|\la^{\prime}_{12}|^2+|\la^{\prime}_{11}|^2}
 \ee
 Numerically, taking $\langle\rho \rangle  =v \simeq v_{electroweak}$ = 246 GeV, we   obtain
 \bea
  \epsilon &\simeq& -2 \times 10^{-6}\left(\fr{M_{R1}}{10^{10}\
  \textrm{GeV}}\right)\left( \fr{m_{\nu 3}}{0.05\
   \textrm{eV}}\right) \delta_{eff}
 \eea
As considered in section \ref{seesaw}, there is no loop
 correction to the Dirac mass matrix  $M_D$; the
effective coupling $\lambda^\prime_{11}=0$ is the same as the
coupling at the tree level.
 Assuming
the coupling $ \la^{\prime}_{12}= |\la | e^ {i \delta_{12}},
\la_{13}^\prime =|\la | e^ {i \delta_{13}}$, we get the  effective
CP-violating phase
 \be
\delta_{eff}=\fr{\sin \delta_{13}+\fr{m_{\nu 2}}{m_{\nu 3} }\sin
\delta_{12}}{2}. \ee As far as we know, the neutrino oscillation
data is given in \cite{Oscillation} as follows: \be
\Delta_{12}^2=7.59 \times 10^{-5} \textrm{eV}^2, \Delta_{13}^2=
2.43 \times 10^{-3} \textrm{eV}^2 \ee Assuming that the neutrino
mass spectrum has a  normal hierarchy, \be M_v= Diag\left(m_o,
\sqrt{m_o^2+\Delta_{12}^2}, \sqrt{m_o^2+\Delta_{13}^2}\right), \ee
leads to the product of the maximal CP asymmetry and the heaviest
light neutrino mass, which is presented in  Fig. \ref{specsugrah}.

 On the other hand, the ratio of the lepton number
 to entropy density after preheating
can be written as \be \fr{n_B}{s}\simeq 10^{-10}B_r(\phi
\rightarrow  \nu_{R 1}\nu_{R1})\left(\fr{T_R}{10^{6}\
\textrm{GeV}}\right)\left(\fr{M_{R1}}{M_\phi}\right)\left(
\fr{\delta_{eff}m_{\nu 3}}{0.05\  \textrm{eV}}\right). \ee The
cosmological constraint on the gravitino abundance gives a bound
on the reheating temperature \cite{Particle}: $T_R < 10^{7}$ GeV.
Assuming that  the reheating temperature is $T_R= 10^6$ GeV and
combining with the observed baryon number to entropy ratio, we get
a constraint on the heaviest light neutrino as \be m_{\nu 3}
> 0.01\ \textrm{eV}. \ee Taking the maximal CP violating phases
from Fig.\ref{specsugrah}, we can roughly estimate the value
$\delta_{eff}m_{\nu 3}= 0.05$. Hence, in order to satisfy the
observed value of the baryon asymmetry \cite{Particle} \be Y_B
=\fr{n_B}{s} = 0.87 \times 10^{-10}, \ee
 the ratio $\fr{M_{R1}}{M_{\phi}}$ must  satisfy
 \be
 \fr{M_{R1}}{M_{\phi}} = 0.87.
 \ee
If we combine  the cosmological constraint on the gravitino
abundance
 ($T_R < 10^{7}$ GeV) with  Eqs. (\ref{Gam}) and (\ref{Aeff}),  we obtain
the constraint on the effective coupling
 \be
 A_{eff} < \fr{10^{-\fr{3}{2}}}{M_{\phi}^{\fr{1}{2}}}.
 \label{est}\ee
From  Eq.(\ref{Aeff}) and  Eq.(\ref{est}),  the constraint on the
inflaton mass is given by \be M_\phi^{\fr{3}{2}}\leq 2 \times
\fr{10^{\fr{-7}{2}}}{\al^2} \times g^{\prime 2}w^2. \ee Taking
into account $g^{\prime } w \propto 10^3$ GeV, we get  the
constraint on the inflaton mass: \be M_{\phi}^{\fr{3}{2}} \leq
2\times \fr{10^{\fr{5}{2}}}{\al^2}. \ee Note that the constraint
on the coupling $\al$ has been given in \cite{LHIn}, namely the
value of coupling $\al$ should be smaller than $10^{-4}$. The
inflaton mass is roughly estimated in Table \ref{table}.

\begin{table}[h]
\caption{
    Coupling constant $\al$ and inflaton mass}
\begin{center}
\begin{tabular}{|c|c|c|c|c|c|c|c|}
  \hline
  $\al$ & $ 10^{-4}$&$ 10^{-5}$ & $10^{-6} $& $10^{-7}$ & $10^{-8}$ & $10^{-9}$ & $10^{-10}$ \\
  \hline
 $ M_\phi^{max}$[ GeV] & $2\times 10^7$ & $4 \times 10^8$&$ 9.2 \times 10^9 $&
  $2\times 10^{11}$ & $4 \times 10^{12}$& $9.2 \times 10^{13}$ & $2\times 10^{15}$ \\
  \hline
\end{tabular}
 \label{table}
\end{center}
\end{table}

 Table \ref{table} shows  that the constraint on the coupling
$\al$ as $(\al \in [10^{-4}, 10^{-10}])$
 leads to the inflaton mass around $M_\phi \in [10^7, 10^{15}]$
GeV. These values not only produce the observed value of the
baryon asymmetry but also are suitable to the hybrid inflationary
scenario given in   \cite{LHIn}.

In short, non-thermal leptongenesis scenario via inflaton decay to
the pair of right handed neutrinos is forbidden at the tree level.
However, this process is available  at the one-loop level. By
taking the reheating temperature $T_R= 10^6$ GeV,  we can solve
the gravitino problem.  Due to  $\delta_{eff}<1$, the heavies
light neutrinos mass satisfies both
 the cosmological
 constraints
 and the oscillation data $m_{\nu 3} \simeq
 \frac {0.05}{\delta_{eff}} $ eV.  We have  obtained the constraint
on the lightest heavy right-handed neutrino: its mass  is smaller
than those of inflaton, namely $ \fr{M_{R1}}{M_{\phi}} =0.87$. It
is worth noting that
 the cosmological constraint on the gravitino abundance gives a bound on the Higgs-self
couplings  and inflation  mass,  which naturally fit to  our
inflation scenario.

\section{Summary and conclusions}
\label{conclusion}

In this  paper, non-thermal leptogenesis in which the heavy
Majorana neutrinos are produced through inflaton decays in the
supersymmetric economical 3-3-1 model with the inflationary
scenario has been considered.

We have shown that the problem in the supersymmetric economical
3-3-1 model (without the inflationary scenario) is  the same as in
the non-supersymmetric version: neutrino masses are unrealistic:
there is no larger hierarchy between $M_{L}, M_R$ and $M_D$.  It
is difficult to obtain the seesaw mechanism  in this scenario.

Fortunately, in the model with inflationary scenario, the
lepton-number-violating interactions among the inflaton and
right-handed neutrinos appear at the one-loop level. Thus, it not
only gives a solution for the above puzzle but also gives a chance
for studying non-thermal leptogenesis scenario.

Our analysis has shown that the leptogenesis works without
overproduction of gravitinos if reheating temperature $T_R= 10^6$
GeV and the lightest  heavy   right-handed neutrino  mass
satisfies $ M_{R1} =\fr{ M_{\phi}}{0.87} $. This result satisfies
also the cosmological constraint $m_{\nu 3} \simeq
\frac{0.05}{\delta_{eff}} $ eV with $\delta_{eff} <1$.

One of the criteria for the inflationary scenario, beside
providing the predictions in good  agreement with observations of
the microwave background and large scale structure formation,  is
an explanation of the origin of the observed baryon asymmetry. For
this aim,  we note that the model under consideration contains the
lepton-number-violating interactions among the inflation and the
right-handed neutrinos at one-loop  level, and this is a reason
for the successful leptogenesis scenario considered in this work.

\section*{Acknowledgement}

D. T. H. is grateful to Nishina Fellowship Foundation for
financial support. She would like to thank Prof. Y. Okada and
Members of Theory Group at KEK for support and comments. H. N. L.
would like to thank Palash B. Pal and   Theory
 Division, Saha Institute of Nuclear Physics for hospitality and
  financial support of his visit where this work was  completed.
 This work was supported in part by the National Foundation for Science
and Technology Development (NAFOSTED)  under grant  No:
103.01.16.09.

\appendix
\section{\label{ap1} Feynman integration}
In this Appendix, we present evaluation of the integral. \be
I_1(a,b,c)=\int\fr{d^4p}{(2\pi)^4}\fr{1}{(p^2-a)^2(p^2-b)(p^2-c)},
\ee
  \bea
I(a,b,c)&\equiv&
\int\fr{d^4p}{(2\pi)^4}\fr{p^2}{(p^2-a)^2(p^2-b)(p^2-c)},\label{tichphan}\eea
where $a,b,c>0$ and $I(a,b,c)=I(a,c,b)$ should be noted in use.
\bea I_1(a,b,c)=\fr{-i}{16\pi^2}\left\{\fr{a\ln
a}{(a-b)(a-c)}+\fr{b\ln b}{(b-a)(b-c)}+\fr{c\ln
c}{(c-b)(c-a)}\right\}.\label{interg}\eea

\bea I(a,b,c)&=&
\int\fr{d^4p}{(2\pi)^4}\left[\fr{1}{(p^2-a)(p^2-b)(p^2-c)} +
\fr{a}{(p^2-a)^2(p^2-b)(p^2-c)}\right]\crn&=&
\fr{-i}{16\pi^2}\left\{\fr{a(2\ln a +1)}{(a-b)(a-c)} -\fr{a^2(2a
-b -c)\ln a }{(a-b)^2(a-c)^2}\right.\crn && + \left\{\fr{b^2\ln
b}{(b-a)^2(b-c)}+\fr{c^2\ln c}{(c-a)^2(c-b)}\right\}.\nn\eea

If $a,b\gg c$ or $c\simeq 0 $, we have an approximation as follows
\be I(a,b,c)\simeq
-\fr{i}{16\pi^2}\fr{1}{a-b}\left[1-\fr{b}{a-b}\ln\fr{a}{b}\right].
\label{ketqua}\ee

In the other case with $b=c$ and $b\neq a$, we have also \be
I(a,b)\equiv I(a,b,b)=-\fr{i}{16
\pi^2}\left[\fr{a+b}{(a-b)^2}-\fr{2ab}{(a-b)^3}\ln\fr{a}{b}\right],\label{ap4}\ee
where, also, $I(a,b)=I(b,a)$ should be noted in use.

If $b\gg a$ or $a\simeq 0$, we have the following approximation
\be I(a,b)\simeq-\fr{i}{16\pi^2b}.\label{ap3}\ee

Let us note that the above approximations $aI(a,b,c)$ (or
$bI(a,b,c)$) and $bI(a,b)$ are kept in the orders up to ${\cal
O}(c/a,c/b)$ and ${\cal O}(a/b)$, respectively.

\end{document}